\documentclass[a4paper,12pt]{article}
\pdfoutput=1
\usepackage[margin=1in]{geometry}
\usepackage{graphicx,xcolor}
\usepackage{mathtools,braket,blkarray}
\allowdisplaybreaks
\usepackage{amssymb,amsfonts,bbm,relsize}
\usepackage[width=0.9\textwidth,footnotesize]{caption}
\usepackage[font=scriptsize]{subcaption}
\usepackage{cite,hyperref,url}
\hypersetup{pageanchor=false}
\usepackage{booktabs,multirow,makecell}
\usepackage[utf8]{inputenc}

\newcommand{\overbar}[1]{\mkern 1.5mu\overline{\mkern-1.5mu#1\mkern-1.5mu}\mkern 1.5mu}
\newcommand{\pmatr}[1]{\begin{pmatrix} #1 \end{pmatrix}}

\newcommand{\diag}{\mathrm{diag}}

\begin{document}

\begin{titlepage}
\begin{center}
{\bf\Large ${\bf SU(3)\times SO(10)}$ in 6d} \\[12mm]
Francisco~J.~de~Anda$^{\dagger}$%
\footnote{E-mail: \texttt{franciscojosedea@gmail.com}},
Stephen~F.~King$^{\star}$%
\footnote{E-mail: \texttt{king@soton.ac.uk}}
\\[-2mm]

\end{center}
\vspace*{0.50cm}
\centerline{$^{\star}$ \it
School of Physics and Astronomy, University of Southampton,}
\centerline{\it
SO17 1BJ Southampton, United Kingdom }
\vspace*{0.2cm}
\centerline{$^{\dagger}$ \it
Tepatitl{\'a}n's Institute for Theoretical Studies, C.P. 47600, Jalisco, M{\'e}xico}
\vspace*{1.20cm}

\begin{abstract}
{\noindent We discuss a simple and elegant $SU(3)\times SO(10)$ family unified 
gauge theory in 6d compactified on a torus with the orbifold $T_2/Z_2^3$ and supplemented by a $Z_6\times Z_3$ discrete symmetry.
The orbifold boundary conditions generate all the desired $SU(3)$ breaking vacuum alignments,
including the $(0,1,-1)$ and $(1,3,-1)$ alignments 
of the Littlest Seesaw model for
atmospheric and solar neutrino mixing, as well as the usual $SO(10)$ breaking with doublet-triplet splitting. 
The absence of driving and messenger fields considerably simplifies the field content of the model.
It naturally explains why there are three families of quarks and leptons, and 
accounts for all their
masses, mixing angles and CP phases via rather elegant looking
Yukawa and Majorana matrices in the theory basis. The resulting model controls proton decay and allows successful Leptogenesis.}
\end{abstract}
\end{titlepage}

\section{Introduction}

The Standard Model (SM) does not explain the existence of either the three families of quarks and leptons or the three gauge forces.
The quest for unification of the three forces led to the original proposal of $SU(5)$ Grand Unified Theory (GUT) \cite{Georgi:1974sy}, while 
the discovery of neutrino mass and mixing motivates an $SO(10)$ GUT.
Gauge coupling unification (in a single step) provides the traditional motivation for 
TeV scale supersymmetry (SUSY) \cite{Langacker:1980js,Chung:2003fi}, which, although so far elusive at the LHC, may yet eventually be discovered 
in the future. 

The explanation of the three families of quarks and leptons is less clear and there have been various proposals put forwards.
One idea is to extend the GUT symmetry to large groups which can accommodate three families such as $SU(8)$ or $O(16)$ \cite{su8}.
Another approach is to introduce a commuting family symmetry such as $SU(3)$ or one of its subgroups.
If the three families of quarks and leptons are unified into a triplet of an $SU(3)$ gauged family symmetry, this could 
provide a reason for the origin of the three families which would be analogous to the three colours of quarks in QCD.

Attempts have been made to formulate an $SU(3)\times SO(10)$ gauge theory which includes the unification of the three families 
and the three gauge forces, although not within a single gauge group~\cite{King:2001uz}.
Similar models with discrete subgroups of $SU(3)$
such as $A_4$ or $S_4$ \cite{Ishimori:2010au} have also been combined with GUTs \cite{King:2017guk,Hagedorn:2010th,Antusch:2014poa,Bjorkeroth:2015ora}. 
The problem is that the most ambitious such 
complete theories also require additional sectors to achieve the desired vacuum alignments and to 
break the gauge symmetry to the SM with doublet-triplet Higgs splitting, leading to somewhat involved models 
\cite{Antusch:2014poa,Bjorkeroth:2015ora,Bjorkeroth:2017ybg,deAnda:2017yeb}.

There is a top-down motivation for considering such models coming from string theory formulated 
in extra dimensions. For example $E_8\times E_8$ heterotic string theory can accommodate $SU(3)\times SO(10)$.
Many of the complications of doublet-triplet splitting are avoided by assuming the existence of extra dimensions \cite{Kawamura:2000ev}.
For example, extra dimensional models have been constructed based on combining 
$A_4$ or $S_4$ with $SU(5)$ \cite{Altarelli:2008bg,Burrows:2009pi,Burrows:2010wz,deAnda:2018oik}.
In such theories, the discrete Family Symmetry could have a dynamical origin as a result
of the compactification of a 6d theory down to 4d
\cite{Altarelli:2006kg,Adulpravitchai:2010na,Adulpravitchai:2009id,Asaka:2001eh,Burrows:2009pi,Burrows:2010wz,
deAnda:2018oik}. The connection of such orbifold
compactifications to string theory has been discussed in \cite{Kobayashi:2006wq}.

In this paper we discuss a simple and elegant $SU(3)\times SO(10)$ family unified 
gauge theory in 6d compactified on a torus with the orbifold $T^2/Z_2^3$ and supplemented by a $Z_6\times Z_3$ discrete symmetry.
The orbifold boundary conditions generate all the desired $SU(3)$ breaking vacuum expectation values (VEVs)
including the $(0,1,-1)$ and $(1,3,-1)$ vacuum alignments (CSD3)
of the Littlest Seesaw model \cite{King:2013iva,King:2015dvf,Ballett:2016yod} for
atmospheric and solar neutrino mixing, as well as the usual $SO(10)$ breaking with doublet-triplet splitting. 
The absence of driving and messenger fields considerably simplifies the field content of the model.
It naturally explains why there are three families of quarks and leptons, and 
accounts for all their
masses, mixing angles and CP phases via rather elegant looking
Yukawa and Majorana matrices in the theory basis. 
The resulting model controls proton decay and allows successful Leptogenesis.

The layout of the remainder of the paper is as follows. In sec. \ref{sec:orb} we present the details of the orbifold and boundary conditions. In sec. \ref{sec:ficont} we shoe the full field content of the model and how it behaves in the extra dimensions. In sec. \ref{sec:flavon} we show how the $SU(3)$ breaking vacuum alignments 
are fixed in through boundary conditions. In sec. \ref{sec:yuk}, the effective Yukawa terms of the model, the fermion mass matrices and a numerical fit are presented. In sec. \ref{sec:protdec} we show how proton decay is controlled.  In sec. \ref{sec:lepto} we show how the Baryon Asymmetry of the Universe (BAU) can be obtained through Leptogenesis in our model. Sect. \ref{conclusion} concludes the paper.
In Appendix~\ref{app:csd2} we discuss the implications of an alternative $SU(3)$ breaking vacuum alignment $(1,0,2)$ (CSD2)
\cite{Antusch:2011ic}.

\section{Orbifolding}
\label{sec:orb}
We assume as gauge symmetry $SU(3) \times SO(10)$. We also assume that the spacetime is the 6d manifold  $\mathcal{M}=R^4\times T^2$, where the torus is defined by 
\begin{equation}\begin{split}
(x^5,x^6)&=(x^5+2\pi R_1,x^6),
\\ (x^5,x^6)&=(x^5,x^6+2\pi R_2).
\label{eq:tras}
\end{split}\end{equation}
We will use interchangeably the complex notation
\begin{equation}
z=x_5+i x_6,
\end{equation}
where, for simplicity in this notation, we will absorb the dimension so that $R_1=R_2=1$. 

The extra dimensions are actually orbifolded so that they are $T_2/ Z_2^{3}$.
The orbifolding leaves 4 invariant 4d branes
\begin{equation}
z_i=\ 0, \ \frac{1}{2},\ \frac{i}{2},\ \frac{1+i}{2}.
\end{equation}
We locate one $\mathbb{Z}_2$ boundary condition on the branes
\begin{equation}
\mathbb{Z}_2:\ \ \ \tilde{z}_i=-\tilde{z}_i, \ \ \ {\rm where} \ \ \ \tilde{z}_i=z+z_i,
\label{eq:par}
\end{equation}
where each boundary condition is defined by a matrix $P_i$ that satisfies $P_i^2=I$ and $i=0,1/2,i/2$. We aim that these boundary conditions break the gauge symmetry into the MSSM.

The boundary conditions are chosen to be
\begin{equation}\begin{split}
P_0&=\mathbb{I}_{10} \otimes SU,\\
P_{1/2}&=P_{PS}\otimes T_{13},\\
P_{i/2}&=P_{GG}\otimes T_1,
\end{split}\end{equation}
where
\begin{equation}
P_{GG}=diag(1,1,1,1,1)\otimes \sigma_2,\ \ \ P_{PS}=diag(-1,-1,-1,1,1)\otimes \sigma_0,
\label{eq:ggps}
\end{equation}
and
\begin{equation}\begin{split}
SU=\frac{1}{3}\left(\begin{array}{ccc} -1 & 2 & 2 \\ 2 & 2 & -1 \\ 2 & -1 & 2 \end{array}\right), \ \ \ T_1=\left(\begin{array}{ccc} -1 & 0 & 0 \\ 0 & 1 & 0 \\ 0 & 0 & 1 \end{array}\right), \ \ \ T_{13}=\left(\begin{array}{ccc} 0 & 0 & -1 \\ 0 & 1 & 0 \\ -1 & 0 & 0 \end{array}\right).
\end{split}\end{equation}

The above $P_{GG}$ boundary condition, when applied to an $SO(10)$ adjoint, 
breaks the gauge group to $SU(5)\times U(1)_X$. To see this explicitly, we may write the adjoint, which is a $10\times 10$ antisymmetric real matrix (with 45 components),
in terms of $5\times 5$ sub-matrices which transform as 
\begin{equation}
P_{GG}\left(\begin{array}{cc} A & A_5 \\ -A_5^\dagger & A' \end{array}\right)P_{GG}=\left(\begin{array}{cc} A' & A_5^\dagger \\ -A_5 & A \end{array}\right),
\end{equation}
where we may see that the $A_5$ submatrix is preserved, provided it is hermitian. This is a  $5\times 5$ submatrix, with 25 components, that correspond to the generators of $SU(5)\times U(1)_X$.

The above boundary condition $P_{PS}$, when applied to an $SO(10)$ adjoint, 
breaks the gauge group to the Pati-Salam gauge group.
To see this, we rotate to an equivalent basis via a matrix $R$ that satisfies $R^2=1$ so that
\begin{equation}
R\ P_{PS}\ R=R\ \diag(-\mathbb{I}_{3\times 3},\mathbb{I}_{2\times 2},-\mathbb{I}_{3\times 3},\mathbb{I}_{2\times 2})\ R=diag(-\mathbb{I}_{6\times 6},\mathbb{I}_{4\times 4}),
\end{equation}
Then we may write the $SO(10)$ adjoint in terms of $6\times 6 $ and $4\times 4 $ matrices which transform as 
\begin{equation}
P_{PS}\left(\begin{array}{cc} A_{6\times 6} & A_{6\times 4} \\ -A_{6\times 4}^\dagger &A _{4\times 4} \end{array}\right)P_{PS}=\left(\begin{array}{cc} A_{6\times 6} & -A_{6\times 4} \\ A_{6\times 4}^\dagger &A _{4\times 4} \end{array}\right),
\end{equation}
so that the antisymmetric real matrices $A_{6\times 6} ,A_{4\times 4} $ are preserved. These matrices generate $SO(6)\times SO(4)$ which is isomorphic to the Pati-Salam group.

To summarise, each boundary condition breaks the symmetry \cite{Adulpravitchai:2010na}
\begin{equation}\begin{split}
P_{GG}&:\ \ \ SO(10)\to SU(5)\times U(1)_X,\\
P_{PS}&:\ \ \ SO(10)\to SU(4)\times SU(2)_L\times SU(2)_R,\\
T_1&:\ \ \ SU(3)_F\to SU(2)_F\times U(1)_F,\\
T_{13}&:\ \ \ SU(3)_F\to U(1)_F\times U(1)_{F'},\\
SU&:\ \ \ SU(3)_F\to \mathbb{Z}_2,\\
\mathbb{I}&:\ \ \ \mathcal{N}=2\to\mathcal{N}=1 \ SUSY.
\end{split}\end{equation}
Together they break $SO(10)\times SU(3)_F\to SU(3)_C\times SU(2)_L\times U(1)_Y\times U(1)_X$ with simple SUSY. The flavour symmetry $SU(3)_F$ is completely broken.

\section{Field content}
\label{sec:ficont}

The field content of the model is listed in table \ref{tab:funfields}. They contain the SM fermions, Higgses, flavons and GUT breaking fields. We remark that only spinorial, fundamental and adjoint representations are used. The field content is rather simple, especially 
when compared to 4d models which aim to be as complete as this one
\cite{Bjorkeroth:2015ora, Bjorkeroth:2015uou, Bjorkeroth:2017ybg,deAnda:2017yeb,deAnda:2018oik}, due to the absence of driving and messenger fields
in the present model. Note that, in addition, there may be other spectator fields (not shown) 
which play no part in the model construction but are there to cancel anomalies,
e.g. additional $3$ representations which any full string theory construction would automatically provide. We assume that they do not obtain any VEV so they do not affect the Yukawa structure.

\begin{table}[ht]
\centering
	\begin{tabular}[t]{| c | c@{\hskip 5pt}c | c c| c c c |}
\hline
\multirow{2}{*}{\rule{0pt}{4ex}Field}	& \multicolumn{4}{c |}{Representation} & \multicolumn{3}{c |}{Localization}\\
\cline{2-8}
\rule{0pt}{3ex}			& $SU(3)$ & $SO(10)$ &  $ Z_
6$&$Z_3$ &$P_{0}$ & $P_{1/2}$ & $P_{i/2}$ \\ [0.75ex]
\hline \hline
\rule{0pt}{3ex}%
$\psi$ 			& $\bar{3}$ & 16 & 0 & 0& &   &  \\
\rule{0pt}{3ex}%
$H_{10}^{u}$ & 1 & 10 & 0 & 0 &+1 &+1&+1\\
$H_{10}^{d}$ & 1 & 10 & 2 & 0 & +1&+1&-1\\
\rule{0pt}{3ex}%
$H_{\overbar{16}}$ & 1 & $\overbar{16}$ & 0 & 0 & +1&+1  &-1 \\
$H_{16}$ & 1 & $16$ & 0 & 0 & +1&+1  &-1 \\
$H_{45}^{X,Y}$ & 1 & $45$ & 0 & 1 & +1 & +1 & +1 \\
$H_{45}^{W,Z}$ & 1 & $45$ & 2 & 1 & +1 & +1 & +1 \\
\rule{0pt}{3ex}%
$\phi_1$ & 3 & 1 & 2 & 1 &+1 & +1&   \\
$\phi_2$ & 3 & 1 & 0 & 1 & +1 &   & +1 \\
$\phi_3$ & 3 & 1 & 3 & 1 &  & +1 & +1 \\
\hline
\end{tabular}
\caption{The simple field content used in constructing the model, including matter, Higgs and flavon superfields.}
\label{tab:funfields}
\end{table}

The superfield $\psi$ contains all the SM fermions. We choose it to be located equally on all of the 4d branes,
\begin{equation}
\psi(x,z)=\psi(x)\Big(\delta^2(z)+\delta^2(z-1/2)+\delta^2(z-i/2)+\delta^2(z-1/2-i/2)\Big),
\end{equation}
which is consistent with the remnant $D_4$ symmetry of the orbifold \cite{Adulpravitchai:2009id}, as well as the underlying $SU(3)$ gauge symmetry.

This is the only field fixed on the 4d branes, and as a consequence is not subject to any boundary conditions.
On the other hand, the flavons are constrained to lie on different 5d branes so that they comply with different boundary conditions. The localization mechanism lies beyond the scope of this work and we treat it as a phenomenological ansatz. This is explained further in section \ref{sec:flavon}.

\subsection{Bulk Superfieds}

All the fields labeled as $H$ obtain a VEV and propagate through the bulk. They are flavour singlet so that they only feel the $P_{GG}, \ P_{PS}$ boundary conditions.

The $H_{u,d}$ are $SO(10)$ fundamentals. They have both positive parity under the condition $P_{PS}$. As can be seen from eq. \ref{eq:ggps}, this condition projects out the triplets, solving the doublet-triplet splitting and leaving only the two doublets in each one. They have opposite parities under the $P_{GG}$ which breaks $SO(10)$ into $SU(5)\times U(1)$. The positive parity projects out the $\overbar{5}$ inside the $10$, while the negative parity projects out the $5$. With both conditions, only one doublet is left massless inside each $H_{u,d}$, which would be the MSSM $h_{u,d}$ respectively. There are no more light doublets which allows for standard gauge coupling unification.

We assume that the $H_{\overbar{16}}$ field develops 
a GUT scale VEV in the singlet $N$ direction in order to break the $U(1)_X$ gauge group, which survives after the rank preserving orbifolding,
and hence allow RHN Majorana masses. This assumption is at least consistent since 
the $H_{\overbar{16}}$ propagates through the bulk and complies with the boundary conditions. 
\footnote{The positive parity under $P_{GG}$ would project out the $\overbar{10},1$ components of the $H_{\overbar{16}}$,
while the negative one would project out the $5$. The positive $P_{PS}$ parity would project out the left fields and the negative parity would project out the right fields. The chosen parities for $H_{\overbar{16}}$ hence leave as possible light modes the corresponding right fields inside the $\overbar{10},1$. These would correspond to the SM $N,u,e$ superfields and the VEV must be aligned with one of them. This can always be rotated to be in the $N$ direction.} The field $H_{16}$, with the same boundary conditions, is included in order to allow renormalizable masses at the GUT scale for
all components of the $H_{\overbar{16}}$.

The $H_{45}$ gets a VEV that supplies the difference between charged leptons and down quarks. It propagates in the bulk and its VEV must comply with the boundary conditions. This reduces the alignment possibilities from 45 to 13. From those 13, the ones in the $SU(3),SU(2)$ generators would break the SM. The VEV can be aligned in a linear combination of the generators $U(1)_X, \ U(1)_Y$. We don't assume any specific choice, other than all the $\braket{H_{45}^{X,Y,W,Z}}$ are different.

We assume that the VEVs $\braket{H_{45}^{X,Y,W,Z}},\braket{H_{\overbar{16}}},\braket{\phi_{1,2,3}}$ are driven radiatively at a large scale $ \lesssim \Lambda\sim M_{GUT}$  \cite{Ibanez:1982fr}.

\section{Flavon alignment}
\label{sec:flavon}

\begin{figure}[h]
\centering
\includegraphics[scale=0.2]{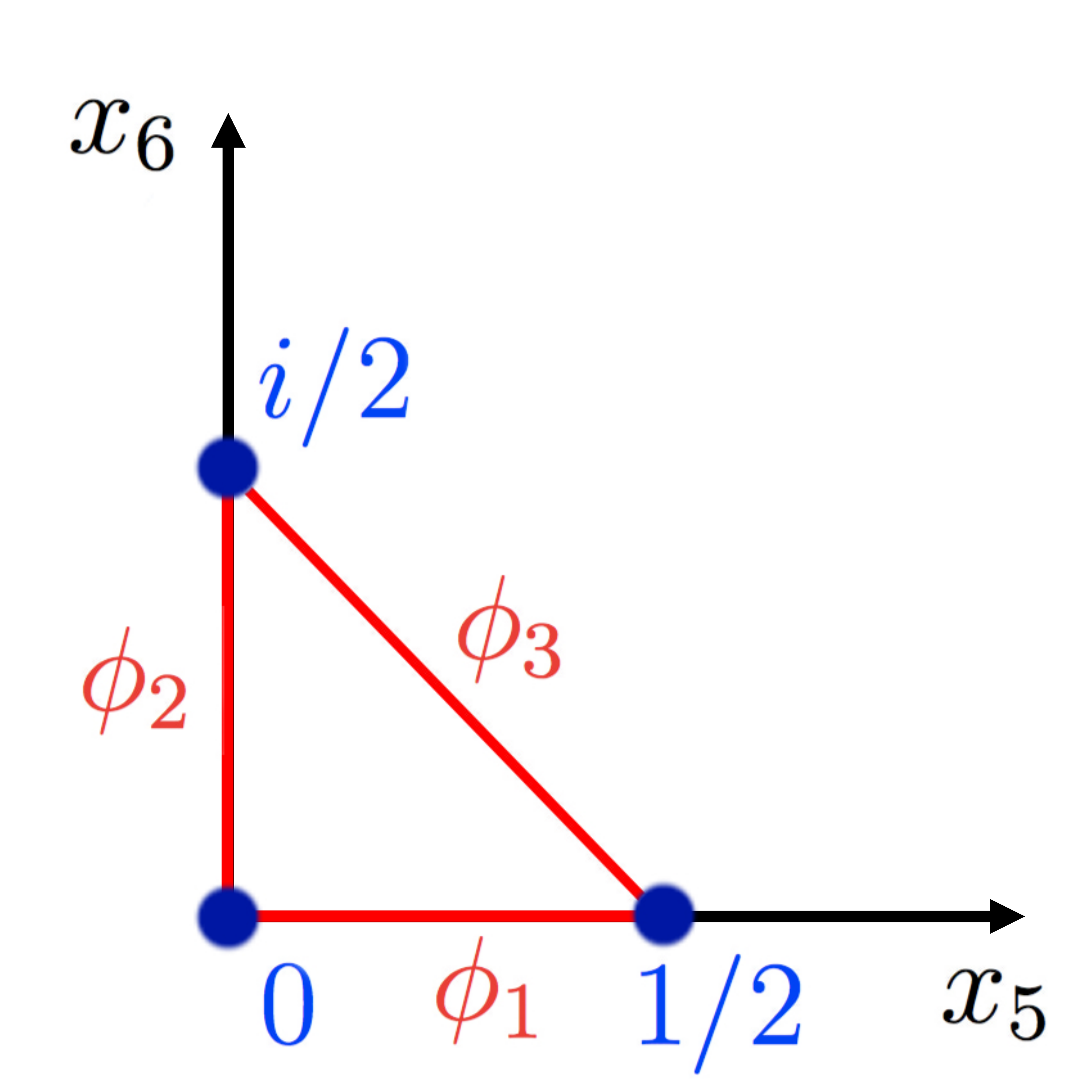}
		\caption{The 5d branes where each flavon propagate. The effective extra dimensional space is the space inside the red triangle. The Flavons propagate through the boundary.}
		\label{fig:5dbrane}
\end{figure}

The model has only 3 flavons that propagate in different 5d branes
\begin{equation}\begin{split}
\phi_1&=\phi_1(x,x_5)\ \delta(x_6),\\
\phi_2&=\phi_2(x,x_6)\ \delta(x_5),\\
\phi_3&=\phi_3(x,x_5)\ \delta(x_5+x_6-1/2),
\end{split}\end{equation}
which can be seen in the figure \ref{fig:5dbrane}. Each flavon propagates in an extra dimensional line and must comply with the boundary conditions
\begin{equation}
\braket{\phi_a}=P_i\braket{\phi_a}.
\end{equation}
These fix completely the flavon VEV alignment.

We obtain the so called CSD3 flavon alignment \cite{King:2016yvg,Bjorkeroth:2015uou,King:2013iva,Bjorkeroth:2017ybg}. This alignment seems to happen more naturally with the discrete flavour symmetry $S_4$ \cite{King:2011zj}.
Inspired by this, we choose one $\mathbb{Z}_2$ boundary condition be the matrix 
\begin{equation}
SU=\frac{1}{3}\left(\begin{array}{ccc} -1 & 2 & 2 \\ 2 & 2 & -1 \\ 2 & -1 & 2 \end{array}\right).
\end{equation}

The flavons $\phi_{1,2}$ must be invariant under the $SU$ matrix, since they have positive parity. This forces their VEVs to be 
\begin{equation}
\braket{\phi_{1,2}}\sim\left(\begin{array}{c}a \\ b \\2a-b\end{array}\right),
\end{equation}
with arbitrary $a,b$.

The VEV $\braket{\phi_1}$ is invariant under $SU$ and $T_{13}$, which forces $b=3a$ and the VEV is aligned as
\begin{equation}
\braket{\phi_{1}} = v_1\left(\begin{array}{c}1 \\ 3 \\-1\end{array}\right).
\end{equation}

The VEV $\braket{\phi_2}$ is invariant under $SU$ and $T_1$, which forces $a=0$ and the VEV is aligned as
\begin{equation}
\braket{\phi_{2}}=v_2\left(\begin{array}{c}0 \\ 1 \\-1\end{array}\right).
\end{equation}

The VEV $\braket{\phi_3}$ is invariant under $T_1$ and $T_{13}$, which forces the first and third entry to vanish, so it is aligned as
\begin{equation}
\braket{\phi_{3}}=v_3\left(\begin{array}{c}0 \\ 1 \\0\end{array}\right).
\end{equation}

This way, all the flavon VEVs are aligned completely through orbifolding, without the need for any superpotential.
The vacuum alignments above are known collectively as CSD3. An alternative vacuum alignment known as
CSD2 is discussed in Appendix~\ref{app:csd2}.

\section{Yukawa terms}
\label{sec:yuk}

In 6d the superpotential must be dimension 5 while a chiral superfield has dimension 2. Any superpotential with interaction terms is non renormalizable, so there is no UV completion adding messenger fields. For this reason we have to consider all order terms.

The effective 4d Yukawa terms allowed by the symmetries are
\begin{equation}
\begin{split}
W_Y &\sim \frac{H_{10}^u(\psi {\phi_1})(\psi{\phi_1})}{\Lambda^3}H_{45}^{X,Y}
+ \frac{H_{10}^u(\psi {\phi_2})(\psi{\phi_2})}{ \Lambda^3}H_{45}^{W,Z}
+ \frac{H_{10}^u(\psi {\phi_3})(\psi{\phi_3})}{ \Lambda^3}H_{45}^{W,Z}
\\ 
&+ \frac{H_{10}^d(\psi {\phi_1})(\psi{\phi_2})}{ \Lambda^3}H_{45}^{W,Z}
+ \frac{H_{10}^d(\psi {\phi_2})(\psi{\phi_2})}{ \Lambda^3}H_{45}^{X,Y}
+ \frac{H_{10}^d(\psi {\phi_3})(\psi{\phi_3})}{\Lambda^3}H_{45}^{X,Y}
\\
&+\frac{H_{\overbar{16}}H_{\overbar{16}} (\psi {\phi_1})(\psi{\phi_1})}{ \Lambda^4}H_{45}^{X,Y}
+ \frac{H_{\overbar{16}}H_{\overbar{16}} (\psi {\phi_2})(\psi{\phi_2})}{\Lambda^4}H_{45}^{W,Z}
+ \frac{H_{\overbar{16}}H_{\overbar{16}} (\psi {\phi_3})(\psi{\phi_3})}{ \Lambda^4}H_{45}^{W,Z}.
\label{eq:efyu}
\end{split}
\end{equation}

This superpotential is responsible for all quark and lepton (including neutrino) masses and mixings.
We shall assume that the flavon VEVs are driven to be hierarchical $v_1\ll v_2\ll v_3$ in order to account for the charged fermion mass hierarchy.

We note that the terms involving $H_u,H_{\overbar{16}}$ each consist of a sum of flavon squared terms. The terms involving $H_d$ have a mixed term $\phi_1\phi_2$. Since we assume $v_1\ll v_2\ll v_3$, 
this will be responsible for the milder hierarchy in the down
sector than the up sector.

All the terms are coupled to two different $H_{45}$ with different dimensionless couplings. The VEVs of the $H_{45}$ treats quarks and leptons differently and we can choose these different couplings to obtain different masses for the charged leptons and down quarks \cite{Bjorkeroth:2015uou}.

The next order terms, with extra flavons are of $O(\phi^8)$, due to the $SU(3)$ symmetry. In the appendix \ref{app:ho} we discuss higher order corrections, with particular focus on those associated with the top quark Yukawa coupling, and also propose a mechanism for naturally suppressing such corrections.

\subsection{Fit friendly matrices}

The fermion mass matrices's structure is determined by the flavons $M_{ij}\sim \sum_{a,b}\braket{\phi_i^a\phi_j^b}$. We make the redefinition
\begin{equation}
\psi\to (\psi_1,\psi_3,-\psi_2)^T.
\label{eq:red}
\end{equation}
We now introduce convenient low energy parameters that effectively come from eq. \ref{eq:efyu}, 
using CSD3 vacuum alignments,
in terms of which the SM fermion mass matrices are \cite{Bjorkeroth:2015uou}
\begin{equation}
\begin{alignedat}{7}
& M^e/ v_d &&=\, & y^e_{1}e^{i\eta_e}\pmatr{0&1&1\\1&2&4\\1&4&6}\,+ \, & y^e_2\pmatr{ 0&0&0 \\ 0&1&1 \\ 0&1&1 }\,+\, &y^e_3e^{i\eta'_e}\pmatr{ 0&0&0 \\ 0&0&0 \\ 0&0&1}\,,  \\
&M^d/v_d&&=\, &y^d_{1}e^{i\eta_d}\pmatr{0&1&1\\1&2&4\\1&4&6}\ +\, &y^d_2\pmatr{ 0&0&0 \\ 0&1&1 \\ 0&1&1 }\,+\, &y^d_3e^{i\eta'_d}\pmatr{ 0&0&0 \\ 0&0&0 \\ 0&0&1}\,,
\\ 
&M^u/v_u&& = \, &y^u_1e^{i\eta_u}\pmatr{1&1&3\\1&1&3\\3&3&9}\ + \, &y^u_2\pmatr{ 0&0&0 \\ 0&1&1 \\ 0&1&1 } + 
&y^u_3e^{i \eta'_u}  \pmatr{ 0&0&0 \\ 0&0&0 \\ 0&0&1 } ,
\\ 
&M_D^\nu/v_u&& = \, &y^\nu_1e^{i\eta_D}\pmatr{1&1&3\\1&1&3\\3&3&9}\ + \, &y^\nu_2\pmatr{ 0&0&0 \\ 0&1&1 \\ 0&1&1 } + &y^\nu_3e^{i \eta'_D}  \pmatr{ 0&0&0 \\ 0&0&0 \\ 0&0&1 } ,
\\ 
& M^\nu_{R} &&= \, &\tilde{M}^\nu_{R1}e^{i\eta_R}\pmatr{1&1&3\\1&1&3\\3&3&9}\ + \, & \tilde{M}^\nu_{R2}\pmatr{ 0&0&0 \\ 0&1&1 \\ 0&1&1 }\, + \, &\tilde{M}^\nu_{R3}e^{i \eta'_R}  \pmatr{ 0&0&0 \\ 0&0&0 \\ 0&0&1 } ,
\\ 
& m^{\nu} &&= \, &\mu_1e^{i\eta_\nu}\pmatr{1&1&3\\1&1&3\\3&3&9}\ + \, & \mu_2\pmatr{ 0&0&0 \\ 0&1&1 \\ 0&1&1 }\, + \, &\mu_3e^{i \eta'_\nu}  \pmatr{ 0&0&0 \\ 0&0&0 \\ 0&0&1 },
\label{eq:masmat}
\end{alignedat}
\end{equation}
where we distinguish the different neutrino mass matrices as $M_D^\nu$ (the Dirac mass matrix), 
$M^\nu_{R}$ (the heavy right-handed neutrino Majorana mass matrix) and $m^{\nu}$ (the light effective left-handed Majorana mass matrix after the seesaw mechanism). Note that the parameters $\tilde{M}^\nu_{Ri}$ (denoted by tildes) differ from the eigenvalues of 
$M^\nu_{R}$ which are later written as ${M}^\nu_{Ri}$ (without tildes). 

We assume that $\braket{H_{45}^{XYWZ}}$ are in general complex so that, together with the $\braket{\phi_i}$  they break CP. Due to the amount of dimensionless constants that couple to each $H_{45}$, we can obtain a free phase in each mass matrix.

The left handed neutrino small masses, $m^\nu$, are generated through the usual seesaw mechanism. Due to the $M^\nu_D,\ M^\nu_R$ mass matrices being rank 1 and with the same structure, the $m^\nu$ has the same structure with  \cite{Bjorkeroth:2015uou,Bjorkeroth:2016lzs}
\begin{equation}
\mu_i=\frac{(y^\nu_i v_u)^2}{\tilde{M}^{\nu}_{Ri}}.
\label{eq:seesaw}
\end{equation}
Furthermore $\eta_\nu=2\eta_D-\eta_R$, which also applies for the primed phases. Note that $\mu_i$ are not equal to the light neutrino mass eigenstates, where the latter are the eigenvalues of the matrix $ m^{\nu}$
written as $m_i$.

At low energies, there are 12 real parameter (9 dimensionless and 3 neutrino masses),  and 8 phases.

We assume the hierarchy between families arise from the flavon VEVs being hierarchical,
for example with the values,
\begin{equation}\begin{split}
v_3\sim \Lambda,\ \ v_2\sim 10^{-1}\Lambda,\ \ v_1\sim 10^{-3}\Lambda,\\
\Lambda\sim M_{GUT}\sim\braket{H_{45}}\sim\braket{H_{\overbar{16}}},
\label{eq:natval}
\end{split}\end{equation}
which would yield the following natural values for the parameters 
\begin{equation}\begin{split}
y^{u,\nu}_1\sim 10^{-6},\ \ y^{e,d}_1\sim 10^{-4},\ \ y_2^{u,\nu,d,e}\sim 10^{-2},\ \ y_{1}^{u,\nu,d,e}\sim 1,\\
\tilde{M}^{\nu}_{R1}\sim 10^{10}\ {\rm GeV},\ \ \tilde{M}^{\nu}_{R2} \sim 10^{14}\ {\rm GeV},\ \ \tilde{M}^{\nu}_{R3}\sim 10^{16}\ {\rm GeV}, \\
{\rm and \  hence} \ \ \ \ \mu_1\sim 1  \ {\rm \mu eV},\ \ \mu_2 \sim 10\ {\rm meV}, \ \ \mu_3 \sim 1\ {\rm eV}.
\label{eq:natyuk}
\end{split}\end{equation}

Concluding, we have 20 low energy parameters to generate the 20 flavour parameters (22 counting Majorana phases).

\subsection{Threshold Corrections}

We have to run the SM Yukawa couplings to the GUT scale, where we do the fit. Since our model does not say anything about SUSY breaking can parametrize its unknown contributions to the running through the threshold corrections \cite{Antusch:2013jca}
\begin{equation}
\begin{split}
y^\mathrm{MSSM}_{u,c,t} 	&\simeq y^\mathrm{SM}_{u,c,t} \csc \beta , \\
y^\mathrm{MSSM}_{d,s} 		&\simeq (1 + \bar{\eta}_q)^{-1} \, y^\mathrm{SM}_{d,s} \sec \beta , \\
y^\mathrm{MSSM}_{b} 		&\simeq (1 + \bar{\eta}_b)^{-1} \, y^\mathrm{SM}_{b} \sec \beta , \\
y^\mathrm{MSSM}_{e,\mu} 	&\simeq (1 + \bar{\eta}_\ell)^{-1} \, y^\mathrm{SM}_{e,\mu} \sec \beta , \\
y^\mathrm{MSSM}_{\tau} 		&\simeq y^\mathrm{MSSM}_{\tau} \sec \beta .
\end{split}
\label{eq:MSSMyukawas}
\end{equation}
The CKM parameters receive the contributions
\begin{equation}
\begin{split}
\theta^{q,\mathrm{MSSM}}_{i3} 	\simeq \frac{1 + \bar{\eta}_b}{1 + \bar{\eta}_q} \, \theta^{q,\mathrm{SM}}_{i3} , \hspace{7mm}
\theta^{q,\mathrm{MSSM}}_{12} 	\simeq \theta^{q,\mathrm{SM}}_{12} , \hspace{7mm}
\delta^{q,\mathrm{MSSM}} 		\simeq \delta^{q,\mathrm{SM}} .
\end{split}
\label{eq:MSSMmixingangles}
\end{equation}
SUSY threshold corrections to the neutrino sector are negligible \cite{Antusch:2013jca,Antusch:2005gp}.

We will be assuming 
\begin{equation}
\tan\beta=10,\ \ \bar{\eta}_b=-0.9,\ \  \bar{\eta}_q=0.4,\ \  \bar{\eta}_l=0,
\end{equation}
since the values $\bar{\eta}_{b,q}$ improve significantly the numerical fit.

The $\eta_b$ parameter is needed to be somewhat large and the leading contributions come from loops either sbottoms and gluinos or stops and higgsinos that add up to \cite{Hall:1993gn}
\begin{equation}
\bar{\eta}_b\simeq \frac{\tan\beta}{16\pi^2}\left(\frac{8}{3}g_3^2\frac{m_{\tilde{g}}\mu}{2m_0^2}+\lambda_t^2\frac{\mu A_t}{m_0^2}\right),
\end{equation}
where $m_0$ denotes the squark masses, $g_3$ the strong coupling, $m_{\tilde{g}}$ the gluino mass and $A_t$ the SUSY softly breaking trilinear coupling involving the stops.
We see that a large contribution can be achieved when
\begin{equation}
m_{\tilde{g}},\mu,A_t >m_0,\hspace{7mm} \tan\beta \gtrsim 10.
\end{equation}
The parameter $\bar\eta_q$ has a similar expression
\begin{equation}
\bar{\eta}_q\simeq \frac{\tan\beta}{16\pi^2}\left(\frac{8}{3}g_3^2\frac{m_{\tilde{g}}\mu}{2m_0^2}\right),
\end{equation}
 but without the contribution of the trilinear coupling. This makes it natural to be smaller but the same order of magnitude.
Finally the parameter
\begin{equation}
\bar{\eta}_l\simeq \frac{\tan\beta}{16\pi^2}\left(\frac{8}{3}g_2^2\frac{m_{\tilde{W}}\mu}{2m_0^2}\right),
\end{equation}
only receives contributions from Winos and Binos and therefore expected to be smaller and negligible.

\subsection{Numerical Fit}

We perform a numerical fit to the flavour observables at the GUT scale. The running of the neutrino parameters is negligible \cite{Antusch:2005gp}.

The numerical fit uses the 12 real parameters, 8 phases and the 2 large threshold corrections, so that we have 22 free real parameters.
As an example, we select $\tan\beta=10$, although a good fit can be obtained with $5<\tan\beta<50$. 
With the setup just mentioned, we can obtain a perfect fit with a $\chi^2\approx 0$. We have 22 flavour parameters at low energies: 6 quark masses, 3 charged lepton masses, 3 light neutrino masses, 4 CKM parameters and 6 PMNS parameters with Majorana masses. We therefore have 22 real parameters in the model for 22 observables (although 3 of them $m_1, \alpha_{21}, \alpha_{31} $ are not yet measured). 

The fit turns out to be quite insensitive to many of the input phases, with the underlying 
CSD3 structure being largely 
responsible for the success of the model as in \cite{Bjorkeroth:2015ora, Bjorkeroth:2015uou, Bjorkeroth:2017ybg,deAnda:2017yeb,deAnda:2018oik}. 
To illustrate this, we consider a benchmark point with 
\begin{equation}
\eta_d=\eta_d'=\eta_e=\eta_e'=0,\ \ \ \eta_u=\eta_u',\ \ \ \eta_\nu=\eta_\nu',
\end{equation}
which reduces the number of input phases to two. 
In addition, motivated by the $Z_6\times Z_3$ symmetry, which could 
play a role in how CP is broken as in \cite{Bjorkeroth:2015ora},
we require that these remaining two phases be a multiple of the 18th roots of unity.
A benchmark point conforming to the above requirements is given in table \ref{tab:fit}. The table also shows a fit related to the alternative 
vacuum alignment discussed in appendix~\ref{app:csd2}.

\begin{table}[!ht]
	\centering
	\footnotesize
	\renewcommand{\arraystretch}{1.1}
	\begin{tabular}{ l cc c cc }
		\toprule
		\multirow{2}{*}{Observable}& \multicolumn{2}{c}{Data} & & \multicolumn{2}{c}{Model best fit} \\
		\cmidrule{2-6}
		& Central value & 1$\sigma$ range  &   & CSD2& CSD3 \\
		\midrule
		$\theta_{12}^\ell$ $/^\circ$ & 33.62 & 32.86 $\to$ 34.38 && 33.75& 31.51 \\ 
		$\theta_{13}^\ell$ $/^\circ$ & 8.54 & 8.57 $\to$ 8.69  && 8.50& 8.54 \\  
		$\theta_{23}^\ell$ $/^\circ$ & 47.20 & 45.30 $\to$ 49.10  && 46.27& 46.85 \\ 
		$\delta^\ell$ $/^\circ$ & 234 & 178 $\to$ 290 && 126& 327  \\
		$y_e$  $/ 10^{-6}$ & 2.05 &  2.03 $\to$ 2.07 && 2.06& 2.06 \\ 
		$y_\mu$  $/ 10^{-4}$ & 4.34 & 4.29 $\to$ 4.39 &&  4.36& 4.36 \\ 
		$y_\tau$  $/ 10^{-3}$ &7.20 & 7.12 $\to$7.28 && 7.23& 7.24 \\ 
		$\Delta m_{21}^2 / (10^{-5} \, \mathrm{eV}^2 ) $ & 7.51  & 7.33 $\to$ 7.69 &&  7.43& 7.39 \\
		$\Delta m_{31}^2 / (10^{-3} \, \mathrm{eV}^2) $ & 2.52  & 2.48 $\to$ 2.56 &&  2.49& 2.49 \\
		$m_1$ /meV & & & & 2.37 & 0.28 \\ 
		$m_2$ /meV & && & 8.94&  8.59 \\ 
		$m_3$ /meV & && & 49.97& 49.95 \\
		$\sum m_i$ /meV & \multicolumn{2}{c}{$<$ 230} &&61.28 & 58.84  \\
		$ \alpha_{21} $ $/^\circ$ & & &&  118& 347\\
		$ \alpha_{31} $  $/^\circ$& & && 286& 129  \\
		$ m_{\beta \beta}$ /meV &  \multicolumn{2}{c}{$<$ 61-165}   && 1.48& 2.02\\    	
		\midrule
		$\theta_{12}^q$ $/^\circ$ &13.03 & 12.98 $\to$ 13.07 && 13.02& 13.02  \\	
		$\theta_{13}^q$ $/^\circ$ &0.22 & 0.21 $\to$ 0.23 && 0.22& 0.23  \\
		$\theta_{23}^q$ $/^\circ$ &2.24& 2.20 $\to$ 2.28 && 2.24& 2.23 \\	
		$\delta^q$ $/^\circ$ & 69.22 & 66.10 $\to$ 72.33  & & 69.45& 72.82 \\
		$y_u$  $/ 10^{-6}$ & 2.81 & 1.96$\to$ 3.65 &&2.83& 2.84  \\	
		$y_c$  $/ 10^{-3}$ & 1.41& 1.40 $\to$ 1.43 && 1.42& 1.42  \\	
		$y_t$  	  & 0.53& 0.49 $\to$ 0.56  && 0.54& 0.52 \\
		$y_d$  $/ 10^{-6}$ & 4.82 & 4.28 $\to$ 5.35 && 5.09& 5.19  \\	
		$y_s$  $/ 10^{-5}$ & 9.65 & 9.16 $\to$ 10.13&& 9.51& 9.65 \\
		$y_b$   $/ 10^{-3}$	  & 5.43 & 5.31 $\to$ 5.54 && 5.44& 5.39  \\
		\bottomrule 
		$\chi^2$ & & & & 4.99& 5.23\\
		\bottomrule
		
	\end{tabular}
	\caption{Flavour observables from experiments compared to the predictions of the model 
	discussed in the main text, based on CSD3 vacuum alignment, as well as for an alternative 
	CSD2 vacuum alignment discussed in Appendix~\ref{app:csd2}.
	The quark masses, charged lepton masses and CKM parameters come from \cite{Antusch:2013jca}. The neutrino observables come from \cite{Esteban:2016qun}. The fits have been performed using the Mixing Parameter Tools (MPT)  package. The SUSY breaking threshold corrections are assumed to be: $\tan\beta=10,\  \bar{\eta}_b=-0.9,\  \bar{\eta}_q=0.4,\   \bar{\eta}_l=0.$}
	\label{tab:fit}
\end{table}

The necessary parameters of the model to obtain this fit are listed in table \ref{tab:parameters}. We can compare these values to the expected natural ones in eq. \ref{eq:natval} and see that all the dimensionless coupling constants have natural values. The $\mu_{1,3}$ are not near from their natural value.

\begin{table}[ht]
	\centering
	\footnotesize
	\renewcommand{\arraystretch}{1.1}
	\begin{tabular}[t]{lr}
		\toprule
		Parameter & Value \\ 
		\midrule
		$y^u_1 \, /10^{-6}$ & $2.84$ \\
		$y^u_2 \, /10^{-2}$ & $0.14$ \\
		$y^u_3$ & $-0.52$ \\
		\rule{0pt}{3ex}%
		$y^d_{1} \, /10^{-4}$ &$-1.55$ \\
		$y^d_2 \, /10^{-3}$ & $-0.32$ \\
		$y^d_3\, $ & $0.54$ \\
		\bottomrule
	
	\end{tabular}
	\hspace*{0.5cm}
	\begin{tabular}[t]{lr}
		\toprule
		Parameter & Value \\ 
		\midrule
		$y^e_{1}$ $/10^{-4}$ & $3.14$ \\
		$y^e_2$ $/10^{-2}$ & $0.41$ \\
		$y^e_3$$/10^{-1}$  & $0.66$ \\
		\rule{0pt}{3ex}%
		$\mu_1 $ /meV & 2.07 \\
		$\mu_2 $ /meV & 31.09 \\
		$\mu_3 $ /meV & 1.89 \\
		\bottomrule
	\end{tabular}
	\hspace*{0.5cm}
	\begin{tabular}[t]{lr}
		\toprule
		Parameter & Value \\ 
		\midrule
		$\eta_u$& $ 7/18 \pi $ \\ 
		$\eta^\prime_u$ & $ 7/18 \pi $ \\
		$\eta_d$, $\eta^\prime_d$& $ 0$ \\ 
		\rule{0pt}{3ex}%
		$\eta_e$, $\eta^\prime_e$& $ 0$ \\ 
		$\eta_\nu$& $ -5/6 \pi $ \\ 
		$\eta^\prime_\nu$ & $ -5/6 \pi $ \\
		\bottomrule
	\end{tabular}
	\caption{Model parameters to generate the fit in table \ref{tab:fit} for CSD3.} 
	\label{tab:parameters}
\end{table}

We showed that we can fit the 22 low energy flavour parameters with 14 real parameters and 8 input phases. 
However, as the benchmark point illustrates, although there are 8 free phases, the results are particularly sensitive to them,
and they may take restricted values such as zero or a particular root of unity, maintaining a good fit to the observables. 

\section{Proton decay}
\label{sec:protdec}

One of the main signatures of GUTs is proton decay. However, it has not been observed and its lifetime is constrained to be \cite{Patrignani:2016xqp}
\begin{equation}
\tau_p>10^{29}\ {\rm yrs}.
\end{equation}

In usual GUTs, the main source for proton decay comes from the new heavy gauge bosons and the color triplets accompanying the Higgs doublets. The triplets are heavy, at the compactification scale, due to the orbifold boundary conditions \cite{Kawamura:2000ev}. So are the extra gauge bosons. We identify the compactification scale with $\Lambda\sim M_{GUT}\sim 2\times 10^{16}\ GeV$, so that the model predicts the proton lifetime to be the same as in usual SO(10) 4d models with \cite{Murayama:1994tc,Langacker:1980js}
\begin{equation}
 \tau_p\sim 10^{29}-10^{30}\ \rm{yrs},
\end{equation}
so that the model barely meets the experimental constraints.

The fact that the compactification scale is so high makes the KK mode contributions to proton decay at least 3 orders of magnitude smaller than the usual sources \cite{Altarelli:2001qj}. These contributions, though small, could eventually provide specific signatures for   extra dimensional GUTs.

There could also be extra contributions to proton decay coming from the extra fields specific to our model. Due to the symmetries of the model the largest contributions would come from the terms
\begin{equation}
\psi\psi\psi\psi\frac{\braket{H_{45}^{W,Z}}^3\braket{H_{45}^{X,Y}}^3}{\Lambda^7_p}, 
\end{equation}
where $\Lambda_p$ is the scale where these term is generated. To comply with the observed proton decay constraints we must have \cite{Murayama:1994tc}
\begin{equation}
\frac{\braket{H_{45}^{W,Z}}^3\braket{H_{45}^{X,Y}}^3M_P^2}{\Lambda^7_p} <3\times 10^{9}\  {\rm GeV}, \ \ \ \Lambda_p> 6\times 10^{17} \ {\rm GeV},
\end{equation}
which is a natural value for this scale. Since this term requires flavour contractions into representations that are not in the original field content, we may expect it to be larger than $\Lambda$.

\section{Leptogenesis}
\label{sec:lepto}

We have seen that the values for $\mu_i$ in this model are not exactly natural. However these quantities appear after the seesaw and relate to $y^\nu,M_R$ as shown in eq. \ref{eq:seesaw}.

If we assume that the heavy RHN mass eigenvalues have the expected natural values, 
of the same order of magnitude as the parameters in eq. \ref{eq:natyuk},
\begin{equation}
M^{\nu}_{R1}\sim 10^{10}\ {\rm GeV},\ \ M^{\nu}_{R2}\sim 10^{14}\ {\rm GeV},\ \ M^{\nu}_{R3}\sim 10^{16}\ {\rm GeV},
\label{eq:natM}
\end{equation}
this would imply that 
\begin{equation}
y^\nu_1\sim 10^{-4},\ \ y^\nu_2\sim 10^{-2},\ \ y^\nu_3\sim 10^{-1},
\label{eq:ynu}
\end{equation}
which is just one order of magnitude away from the natural values for $y^\nu_{3}$ and two orders of magnitude for $y^\nu_{1}$. These values are effectively free in our model and we can tune them to be so without any problem.

Assuming the natural values for $M^{\nu}_{Ri}$ and the deviated ones for $y^\nu$ requires a fine tuning of 1 in 100. However having exactly these values can explain the Baryon Asymmetry of the Universe (BAU) through Leptogenesis.

Leptogenesis generates the BAU through CP violating decay of the lightest RHN into neutrinos generating a lepton asymmetry, then transformed into baryon asymmetry through non perturbative sphaleron processes \cite{DiBari:2012fz}.

Leptogenesis has already been studied with matrices in the CSD3 alignment \cite{Bjorkeroth:2015tsa,King:2015dvf}. The result ultimately depends on the phase $\eta$ which we identify with the leptogenesis phase. With the phase in our fit, to generate the observed BAU the RHN masses must be
\begin{equation}
10^{9}<M^{\nu}_{R1}<10^{11},\ \ \  10^{11}<M^{\nu}_{R2}<10^{13},\ \ \ M^{\nu}_{R3}\sim M_{GUT},
\end{equation}
which are the natural order of magnitude values for the RHN mass parameters as seen in
eq. \ref{eq:natyuk}.
Therefore, if we assume the tuning to obtain the $y^\nu$ as in eq. \ref{eq:ynu}, our model generates the observed BAU through Leptogenesis.

\section{Conclusion}
\label{conclusion}

We have discussed a simple and elegant $SU(3)\times SO(10)$ family unified 
gauge theory in 6d compactified on a torus with the orbifold $T^2/Z_2^3$  and supplemented by a $Z_6\times Z_3$ discrete symmetry.
The orbifold boundary conditions break the symmetry down to $SU(3)_C\times SU(2)_L\times U(1)_Y\times U(1)_X$, achieving 
doublet - triplet splitting and leaving only the light Higgs doublets of the MSSM, with the 
gauge coupling unification scale of order the compactification scale $\Lambda\sim M_{GUT}$.
The $U(1)_X$ is broken by a $H_{\overbar{16}}$ field which develops 
a GUT scale VEV in the singlet $N$ direction, thereby allowing Majorana masses.
Below the GUT scale we then have just the MSSM field content, together with right-handed neutrino masses.

An important and new feature of our model is that  
the orbifold boundary conditions generate all the desired $SU(3)$ breaking vacuum alignments,
such as the $(0,1,-1)$ and $(1,3,-1)$ alignments, without having to introduce an additional superpotential with extra driving fields. 
The absence of driving and messenger fields considerably simplifies the field content of the model which 
requires only twelve superfield multiplets, which is remarkably economical for a complete Flavoured GUT.
Having a gauged $SU(3)$, the model naturally explains why there are three families of quarks and leptons.

The model quantitatively accounts for all quark and lepton (including neutrino) 
masses, mixing angles and CP phases via rather elegant looking
Yukawa and Majorana matrices in the theory basis. 
Although the model involves 14 independent real parameters and 8 phases to fix 22 flavour observables,
we have shown that the successful fit is mainly due to the vacuum alignments, and is insensitive to the precise value of 
many of the phases. To illustrate this we have considered a benchmark point with a restricted set of phases,
and shown that it can achieve a good fit to the observables, with $\chi^2=5$, where most of the real parameters 
take natural $O(1)$ values. However we do not discuss how the large hierarchical flavon VEVs,
responsible for the charged fermion mass hierarchies, are driven.

Finally we remark that the resulting model controls proton decay, with a proton lifetime close to the current limits.
In addition it allows successful Leptogenesis.

\subsection*{Acknowledgements}
We thank Patrick Vadrevange for useful discussions.
SFK acknowledges the STFC Consolidated Grant ST/L000296/1 and the European Union's Horizon 2020 Research and Innovation programme under Marie Sk\l{}odowska-Curie grant agreements Elusives ITN No.\ 674896 and InvisiblesPlus RISE No.\ 690575.

\appendix

\section{Higher order corrections}
\label{app:ho}

In this Appendix we discuss higher order corrections, with particular focus on those associated with the top quark Yukawa coupling, and also propose a mechanism for naturally suppressing such corrections.

Since a superpotential in 6d is always nonrenormalizable, in principle we have to analyze such terms of all orders.
To begin with, the VEVs $\braket{H_{45}}$ are very large and higher powers of them would not be very suppressed. 
However, these VEVs so not affect the matrix structure of the fermion masses. They do affect the relation between charged leptons and down quarks but we can redefine the fit variables so they do not affect at low energies.

Turning to the flavons, the next order terms involving more flavons would be of $O(\phi^8/\Lambda^8)$ due to the $SU(3)$ symmetry. The 
most dangerous such terms involve the VEV $v_3$ which is quite large since it gives the top mass. As we see in the fit, we expect the ratio $v_3/\Lambda\approx 0.7$ so that even large powers of it, like $(v_3/\Lambda)^8\approx 0.05$, are not necessarily negligible.
The largest of these corrections would involve
\begin{equation}
\frac{(\phi_3)^6(\phi_i\phi_j)}{\Lambda^6},
\end{equation}
which is a completely symmetric product into an $SU(3)$ 6 dimensional representation. Since $(\phi_3)^6$ is a singlet under the discrete symmetries, we can have all terms in eq. \ref{eq:efyu} with an extra $(\phi_3)^6/\Lambda^6$. These correct the respective $(\phi_i\phi_j)$ Yukawa terms. They have a suppression of 
$(v_3)^6/\Lambda^6\sim (y^u_3)^3\sim 0.1$. If we choose the corresponding dimensionless coupling constant to be small, less than $0.1$, these terms become negligible.

It is possible to naturally suppress such terms, without appealing to the dimensionless coupling constants,
by adding the messenger-like fields in Table~\ref{tab:mho}.
\begin{table}[ht]
\centering
	\begin{tabular}[t]{| c | c@{\hskip 5pt}c | c c| c c c |}
\hline
\multirow{2}{*}{\rule{0pt}{4ex}Field}	& \multicolumn{4}{c |}{Representation} & \multicolumn{3}{c |}{Localization}\\
\cline{2-8}
\rule{0pt}{3ex}			& $SU(3)$ & $SO(10)$ &  $ Z_
6$&$Z_3$ &$P_{0}$ & $P_{1/2}$ & $P_{i/2}$ \\ [0.75ex]
\hline \hline
\rule{0pt}{3ex}%
$\chi$ & 1 & $ 16$ & 3 & 1 & &&\\
$\bar{\chi}$ & 1 & $\overbar{16}$ & 3 & 2 & &&\\
\hline
\end{tabular}
\caption{Possible extra messenger-like fields fixed on the brane.}
\label{tab:mho}
\end{table}
These fields are located in the same brane as the field $\psi$ so that we have the effective 4d terms after compactification
\begin{equation}
\mathcal{W}_\chi\sim M_\chi \overbar{\chi}\chi+\psi\overbar{\chi}\phi_3+H^u_{10}H_{45}^{X,Y}(\chi\chi+\overbar{\chi}\overbar{\chi}).
\end{equation}
The physical top quark field is then identified as the massless 
linear combination of ${\chi}$ and $\phi_3\psi$. The term $H^u_{10}H_{45}^{X,Y}\chi\chi$ then allows the physical top quark to naturally a have larger mass.
This can also be seen since these effective messengers allow us to make the replacement
\begin{equation}
\frac{\phi_3}{\Lambda}\to\frac{\phi_3}{M_\chi},
\end{equation}
in eq. \ref{eq:efyu}. This allows us to assume $v_3/M_\chi\sim 0.5>>v_3/\Lambda\sim 0.01$, which makes all higher order term negligible.

Any term involving any other flavon or more flavons are completely negligible.

\section{CSD2 vacuum alignment}
\label{app:csd2}

In this appendix we discuss an alternative vacuum alignment (CSD2) \cite{Antusch:2011ic}, 
which can lead to a good fit when combined with $SO(10)$ \cite{deAnda:2017yeb}. 

If we keep all the same model setup but instead of the matrix $T_{13}$ in the condition $P_{1/2}$ we use the matrix
\begin{equation}
T_2=\left(\begin{array}{ccc} 1 & 0 & 0 \\ 0 & -1 & 0 \\ 0 & 0 & 1 \end{array}\right),
\end{equation}
we can obtain the so called CSD2 alignment.

In this case, the VEV $\braket{\phi_1}$ is invariant under $SU$ and $T_{2}$, which forces $b=0$ and the VEV is aligned as
\begin{equation}
\braket{\phi_{1}}\sim\left(\begin{array}{c}1 \\ 0 \\2\end{array}\right).
\end{equation}

The VEV $\braket{\phi_2}$ remains unchanged \begin{equation}
\braket{\phi_{2}}\sim\left(\begin{array}{c}0 \\ 1 \\-1\end{array}\right).
\end{equation}

The VEV $\braket{\phi_3}$ is invariant under $T_1$ and $T_{2}$, which forces the first and second entry to vanish, so it is aligned as
\begin{equation}
\braket{\phi_{3}}\sim\left(\begin{array}{c}0 \\ 0 \\1\end{array}\right).
\end{equation}

These are the 3 VEVs that generate the CSD2 alignment and they have also been achieved through 
orbifold boundary conditions only.

\subsection{CSD2 fit friendly masses}

As we just, with small changes to the model, we can obtain the CSD2 alignment. With this alignment the redefinition in eq. \ref{eq:red} wouldn't need the swap in the last two entries and it would be 
\begin{equation}
\psi\to (\psi_1,-\psi_2,\psi_3)^T
\end{equation}

Using the CSD2 alignment, the SM fermion mass matrices, in terms of the low energy parameters, are
\begin{equation}
\begin{alignedat}{7}
& M^e/ v_d &&=\, & y^e_{1}e^{i\eta_e}\pmatr{0&1&1\\1&0&2\\1&2&4}\,+ \, & y^e_2\pmatr{ 0&0&0 \\ 0&1&1 \\ 0&1&1 }\,+\, &y^e_3e^{i\eta'_e}\pmatr{ 0&0&0 \\ 0&0&0 \\ 0&0&1}\,,  \\
&M^d/v_d&&=\, &y^d_{1}e^{i\eta_d}\pmatr{0&1&1\\1&0&2\\1&2&4}\ +\, &y^d_2\pmatr{ 0&0&0 \\ 0&1&1 \\ 0&1&1 }\,+\, &y^d_3e^{i\eta'_d}\pmatr{ 0&0&0 \\ 0&0&0 \\ 0&0&1}\,,
\\ 
&M^u/v_u&& = \, &y^u_1e^{i\eta_u}\pmatr{1&0&2\\0&0&0\\2&0&4}\ + \, &y^u_2\pmatr{ 0&0&0 \\ 0&1&1 \\ 0&1&1 } + 
&y^u_3e^{i \eta'_u}  \pmatr{ 0&0&0 \\ 0&0&0 \\ 0&0&1 } ,
\\ 
&M_D^\nu/v_u&& = \, &y^\nu_1e^{i\eta_D}\pmatr{1&0&2\\0&0&0\\2&0&4}\ + \, &y^\nu_2\pmatr{ 0&0&0 \\ 0&1&1 \\ 0&1&1 } + &y^\nu_3e^{i \eta'_D}  \pmatr{ 0&0&0 \\ 0&0&0 \\ 0&0&1 } ,
\\ 
& M^{\nu}_R &&= \, &\tilde{M}^\nu_{R1}e^{i\eta_R}\pmatr{1&0&2\\0&0&0\\2&0&4}\ + \, & \tilde{M}^\nu_{R2}\pmatr{ 0&0&0 \\ 0&1&1 \\ 0&1&1 }\, + \, &\tilde{M}^\nu_{R3}e^{i\eta'_R} \pmatr{ 0&0&0 \\ 0&0&0 \\ 0&0&1 } ,
\\ 
&m_\nu && = \, &\mu_1 e^{i\eta_\nu}\pmatr{1&0&2\\0&0&0\\2&0&4}\ + \, &\mu_2\pmatr{ 0&0&0 \\ 0&1&1 \\ 0&1&1 } + &\mu_3e^{i \eta'_\nu}  \pmatr{ 0&0&0 \\ 0&0&0 \\ 0&0&1 } ,
\label{eq:masmat2}
\end{alignedat}
\end{equation}
where all the previous discussion also applies.

\subsection{Numerical fit}

Again, with the CSD2 alignment we can obtain a perfect fit with $\chi^2\approx 0$. However we will also make the arbitrary assumptions to show the predictivity of this setup
\begin{equation}
\eta_d=\eta_d'=\eta_e=\eta_e'=0,\ \ \ \eta_u=\eta_u',
\end{equation}
where we have one less condition than in the CSD3. This brings the amount of physical phases to 3, where one extra phase is needed with respect to CSD3.

In table \ref{tab:fit} we have already shown that we can obtain a good fit in this setup.
The necessary parameters for this fit are shown in the table \ref{tab:parameters2}. 

\begin{table}[ht]
	\centering
	\footnotesize
	\renewcommand{\arraystretch}{1.1}
	\begin{tabular}[t]{lr}
		\toprule
		Parameter & Value \\ 
		\midrule
		$y^u_1 \, /10^{-6}$ & $-2.83$ \\
		$y^u_2 \, /10^{-2}$ &$-0.14$ \\
		$y^u_3$ & $-0.53$ \\
		\rule{0pt}{3ex}%
		$y^d_{1} \, /10^{-4}$ &$-1.53$ \\
		$y^d_2 \, /10^{-3}$ & $-0.63$ \\
		$y^d_3\, $ & $-0.54$ \\
		\bottomrule
	
	\end{tabular}
	\hspace*{0.5cm}
	\begin{tabular}[t]{lr}
		\toprule
		Parameter & Value \\ 
		\midrule
		$y^e_{1}$ $/10^{-4}$ & $2.94$ \\
		$y^e_2$ $/10^{-2}$ & $-0.41$ \\
		$y^e_3$$/10^{-1}$  & $0.75$ \\
		\rule{0pt}{3ex}%
		$\mu_1 $ /meV & 3.43 \\
		$\mu_2 $ /meV & 24.71 \\
		$\mu_3 $ /meV &12.51 \\
		\bottomrule
	\end{tabular}
	\hspace*{0.5cm}
	\begin{tabular}[t]{lr}
		\toprule
		Parameter & Value \\ 
		\midrule
		$\eta_u$& $ 7/18 \pi $ \\ 
		$\eta^\prime_u$ & $ 7/18 \pi $ \\
		$\eta_d$, $\eta^\prime_d$& $ 0$ \\ 
		\rule{0pt}{3ex}%
		$\eta_e$, $\eta^\prime_e$& $ 0 $ \\ 
		$\eta_\nu$& $ 2/9 \pi $ \\ 
		$\eta^\prime_\nu$ & $ 17/18 \pi $ \\
		\bottomrule
	\end{tabular}
	\caption{Model parameters to generate the fit in table \ref{tab:fit} with CSD2.} 
	\label{tab:parameters2}
\end{table}

Again we can see that this setup is as natural as the CSD3 one discussed in the main text.

\end{document}